# jpf-concurrent: An extension of Java PathFinder for java.util.concurrent

Mateusz Ujma
Department of Computer Science,
University of Oxford
Oxford, United Kingdom
Email: mateusz.ujma@cs.ox.ac.uk

Nastaran Shafiei
Department of Computer Science
and Engineering, York University
Toronto, Ontario, Canada
Email: nastaran@cse.yorku.ca

*Abstract*—One of the main challenges when verifying multi-threaded Java applications is the state space explosion problem. Due to thread interleavings, the number of states that the model checker has to verify can grow rapidly and impede the feasibility of verification. In the Java language, the source of thread interleavings can be the system under test as well as the Java Development Kit (JDK) itself. In our paper, we propose a method to minimize the state space explosion problem for applications verified under the Java PathFinder (JPF) model checker. Our method is based on abstracting the state of the application to a smaller domain and implementing application behavior using the Model Java Interface (MJI) of JPF. To show the capabilities of our approach, we have created a JPF extension called jpf-concurrent which abstracts classes from the Java Concurrency Utilities. Several benchmarks proved the usefulness of our approach. In all cases, our implementation was faster than the JDK implementation when running under the JPF model checker. Moreover, our implementation led to significantly smaller state spaces.

## I. INTRODUCTION

The state space explosion problem is known as one of the most challenging issues in model checking [4], [3]. One of the sources of this problem is thread non-determinism. In programs that include thread non-determinism, the concurrent actions can be executed in any order. In such programs, considering all possible interleavings of these concurrent actions can lead to a very large state space that cannot be stored in the available memory resources or processed in reasonable time. It can be shown that the number of program states can increase exponentially with the number of concurrently running components [3].

Java PathFinder (JPF) [10] is an explicit-state model checker that directly works with Java bytecode instructions. The core of JPF is a Java Virtual Machine (JVM) which systematically executes Java programs by exploring all possible thread interleavings. Each execution is a sequence of transitions and each transition is a sequence of bytecode instructions that takes the system from one state to another. While JPF is executing the program, it explores the program state space. JPF, as any other model checker, suffers from the state space explosion problem.

Java applications use the JDK to simplify the development process. For concurrent applications the JDK provides Concurrency Utilities (`java.util.concurrent` package) which contains implementations of basic concurrency constructs. As JPF is mostly used to verify multi-threaded code, a large number of applications verified under JPF use classes from Java Concurrency Utilities. Verifying Java programs which are based on Java Concurrency Utilities gives rise to several challenges. In general, due to thread non-determinism originated from many constructs included in Java Concurrency Utilities, model checking such Java programs leads to very large state spaces. Another challenge is compatibility issues between different JRE/JDK vendors. To improve performance a large number of vendors use private classes to implement `java.util.concurrent` package. These classes tend to contain native methods, the implementation of which is tied to a specific JVM. To run such code JPF would have to handle these native methods for each JVM vendor. This has not been done, and in many cases running code that uses classes from Java Concurrency Utilities triggers an `UnsatisfiedLinkError` error. Finally, unreadable stack traces is another issue which is due to the size of the Java Concurrency Utilities and the number of method calls that are kept on the stack.

To solve the above issues, we have created the jpf-concurrent extension for JPF. In this extension, model classes are created as replacements for the classes included in the Java Concurrency Utilities. The size of the state space is reduced by abstracting the state of the Java Concurrency Utilities classes to a single integer declared within their corresponding model classes. Moreover, the MJI feature of JPF is used to delegate most of the implementation of concurrent classes to the host JVM level where it is not model checked. Using this technique, we keep the complex state of the concurrent objects on the host JVM, and JPF uses an integer to represent different states. Native method compatibility problems have been solved by reimplementing most of the Java Concurrency Utilities classes without using any vendor specific code. Our solution also shortened the size of the stack traces considerably.

The rest of this paper is organized as follows. In Section II we discuss related work and existing solutions. Section III presents an example of how jpf-concurrent can be used for software verification. In Section IV the reader can find detailed description of jpf-concurrent implementation. Section V contains quantitative results of running the benchmarks. Finally, we outline future work in Section VI.

## II. RELATED WORK

There exist several programming interfaces within JPF that can be used to solve the presented issues. To provide implementation for unsupported native methods and reduce the state space JPF contains Model Java Interface (MJI). In resemblance to Java Native Interface (JNI), which is used to transfer the execution from the Java level to the native level, MJI is used to transfer the execution from JPF to the host JVM. All methods that execute on the host JVM using MJI are atomic, therefore, they do not contribute to the state space explosion problem. In the cases where a method has to be executed on the JPF level, in order to avoid thread reschedulings, we can use the class `gov.nasa.jpf.jvm.Verify`. This class includes the methods `beginAtomic()` and `endAtomic()`. All the bytecode instructions surrounded by these two methods are combined into a single transition. Reducing the stack traces can be achieved by means of a listener. Such a listener could extend the `ListenerAdapter` class and implement the `propertyViolated()` method to access and augment the current stack trace. To the best of our knowledge, before jpf-concurrent there was no work that would provide a coherent usage of these techniques that would allow state abstraction on the JPF level and state handling on the host VM level. There are a few other JPF extensions that reimplement JDK libraries to allow verification. jpf-awt [9] focuses on AWT/Swing libraries and uses MJI to reimplement `EventDispatchThread` where it reads UI scripts that are used as drivers for the verification process. Another extension is net-iocache [2], [1] which allows verification of networking applications in the presence of backtracking. Unfortunately, because of unique characteristics of GUI and networking applications none of these techniques can be directly applied to Java Concurrency Utilities.

## III. EXAMPLE

In this section, we use a simple example to explain how the jpf-concurrent extension can be used to verify Java programs. As jpf-concurrent is an extension to JPF, we assume that a fully functional copy of JPF is available. The first step to use our extension is to download it from the jpf-concurrent project website [1]. The next step is to specify a directory where jpf-concurrent can be found in the `site.properties` file. From this point in `*.jpf` files, we can use the `@using` keyword to enable jpf-concurrent for a given target application. The classes modelled within jpf-concurrent follow the same API as their corresponding classes in the Java Concurrency Utilities. Therefore, running JPF with jpf-concurrent enabled does not affect the verification results, i.e., the output produced by JPF with jpf-concurrent will not be different from the output when running it without jpf-concurrent. The only way that using jpf-concurrent affects the verification is the decrease of verification time and support for native calls within classes of Java Concurrency Utilities.

[1] http://babelfish.arc.nasa.gov/trac/jpf/wiki/projects/jpf-concurrent

The content of `ReentrantLockPerformanceTest.jpf` file which is used for running one of the jpf-concurrent benchmarks is shown in Figure 1. Figure 2 contains output of a successful verification.

```
# mode property file for running
JPF on ReentrantLockPerformanceTest example
@using jpf-concurrent
target = ReentrantLockPerformanceTest
```

Fig. 1. ReentrantLockPerformanceTest.jpf

```
./jpf ReentrantLockPerformanceTest.jpf 4
JavaPathfinder v6.0 - (C) RIACS/NASA Ames Research Center
====================================== system under test
application: ReentrantLockPerformanceTest.java
arguments:   4
======================================= search started: 8/21/11 4:55 PM
***** ReentrantLock PERFORMANCE TEST *****
Number of threads = 4
============================================== results
no errors detected
============================================= statistics
elapsed time:      0:00:05
states:            new=18762, visited=36606,
                   backtracked=55367, end=107
search:            maxDepth=35, constraints hit=0
choice generators: thread=18761 (signal=0, lock=2,
                   shared ref=5727), data=0
heap:              new=9818, released=131340,
                   max live=370, gc-cycles=51600
instructions:      769878
max memory:        482MB
loaded code:       classes=82, methods=1382

======================= search finished: 8/21/11 4:55 PM
```

Fig. 2. Running one of jpf-concurrent benchmarks

## IV. TOOL DESCRIPTION

*Java PathFinder Features*

JPF can be considered as a JVM that executes the Java code under test. JPF itself is written in Java, and therefore, it runs on top of another JVM, called the host JVM. Figure 3 shows the different layers involved while JPF is running on target code.

One of the key features of JPF are its model classes. These classes are considered as part of the code under test which is model checked by JPF. Model classes are replacements for Java classes. By implementing a model class corresponding to a certain target class, we can force JPF to not model check the original class, but instead model check the model class as an alternative.

Another key feature of JPF is its MJI. The implementation of MJI is based on classes called native peers. These classes run on top of the host JVM, and they are not model checked by JPF. JPF uses a special name pattern to associate a native peer class to a class of the target code. It also establishes a correspondence between the methods of these classes. Whenever JPF gets to a method that is associated with a corresponding method in the native peer, the execution is transferred to the host JVM. The host JVM executes the body of the native peer method, and after the native peer method returns, JPF continues model checking the rest of the code.

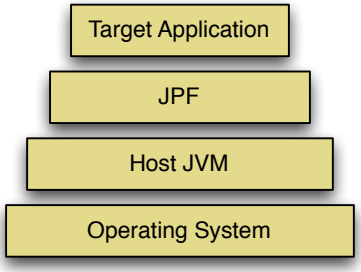

Fig. 3. Different layers involved in JPF execution

*Java Concurrency Utilities*

The Java platform was originally designed to support concurrent programming. Before Java 1.5, the concurrent programming was possible through low-level concurrency primitives (e.g. `synchronized`, `wait`, `notify`) included in the Java language [7]. The Java Concurrency Utilities [5] are included in the Java platform since version 1.5. These utilities add high-level building blocks for creating concurrent classes and applications. Using such well-tested building blocks can avoid potential concurrency bugs such as race conditions, deadlocks, and thread starvation.

Based on the functionality, the Java Concurrency Utilities can be divided into different parts. The following is a description of those parts that have been modelled by the jpf-concurrent project.

- Concurrent Collections: the `java.util.concurrent` package contains a number of concurrent data structure implementations, such as `ConcurrentHashMap` or `ConcurrentLinkedQueue`. Implementations of these collections guarantee that an action that adds an element to the collection has a happens-before relationship [8] with subsequent actions that remove or access that element. Such behavior avoids memory consistency errors.
- Synchronizers: the `java.util.concurrent` package also includes classes that provide commonly-used synchronization idioms, such as `Semaphore` or `Exchanger`.
- Locks: the `java.util.concurrent.locks` package implements a high-performance locking discipline which eliminates some of the limitations of the implicit locks used by synchronized code.
- Atomic Variables: the `java.util.concurrent.atomic` package contains classes that manipulate variables atomically. These classes implement the method `compareAndSet` which provides a conditional atomic update on a specified variable, similar to compare and swap (CAS) [6]. Similar to `volatile` variables, the atomic variables guarantee that a write to the variable has a happens-before relationship with the subsequent read of the same variable.

*Implementation*

In our approach, the major part of the implementation that emulates the behavior of concurrency classes is not model checked, but runs on top of the host JVM. This design is mainly based on the MJI feature of JPF which is used to delegate the calls from Java Concurrency Utilities to the methods that execute at the host JVM level. The actual objects that represent instances of concurrent classes are created and maintained at the host JVM level. The majority of operations on these instances are also performed at the host JVM level.

To reduce the state space of the code, our approach abstracts the states of concurrent objects in the JPF environment, and keeps their actual state in the host JVM. In other words, for each concurrent object there exist two different constructs. One is an abstraction of the object which has a considerably smaller state size, and it is created in the JPF environment. The other represents the actual state of the object, and it is created in the host JVM environment.

To abstract the state of concurrent objects within JPF, our solution introduces model classes that correspond to the Java Concurrency Utilities classes. These model classes have only one private field which is called `version` and is declared as integer variable . From now on, in this paper, we refer to these abstract model classes as JPF-level model classes. Instances of these classes represent the concurrent objects in the JPF environment. Using a single integer to capture a state rather than a complex data structure decreases the state space size considerably.

Consider the class `ConcurrentHashMap` which declares three fields to store the content of a map. By adding a new element to an instance of `ConcurrentHashMap`, the values of these three fields are modified. Moreover, a concurrent access to the values of these fields leads to thread rescheduling and increases the state space. Abstracting these fields with a single integer can considerably decrease thread rescheduling. That does not change the class API as all mentioned fields are private.

In our approach, methods invoked on the concurrent objects in JPF are delegated to the methods in the host JVM that have access to the actual state of these objects. To implement these methods, we created a class in the host JVM level which is completely unknown to JPF. This class is called `Model`. For each concurrent class modelled in jpf-concurrent, there exists a subclass of the class `Model`, which implements the actual operations of the concurrent class. From now on, in this paper, we refer to the subclasses of `Model` as JVM-level model classes. The methods of the JVM-level model classes operate on the actual states of concurrent objects in the host JVM. Therefore, JPF does not associate any transitions to their execution. That reduces the number of interleavings explored by the model checker which results in a smaller state space.

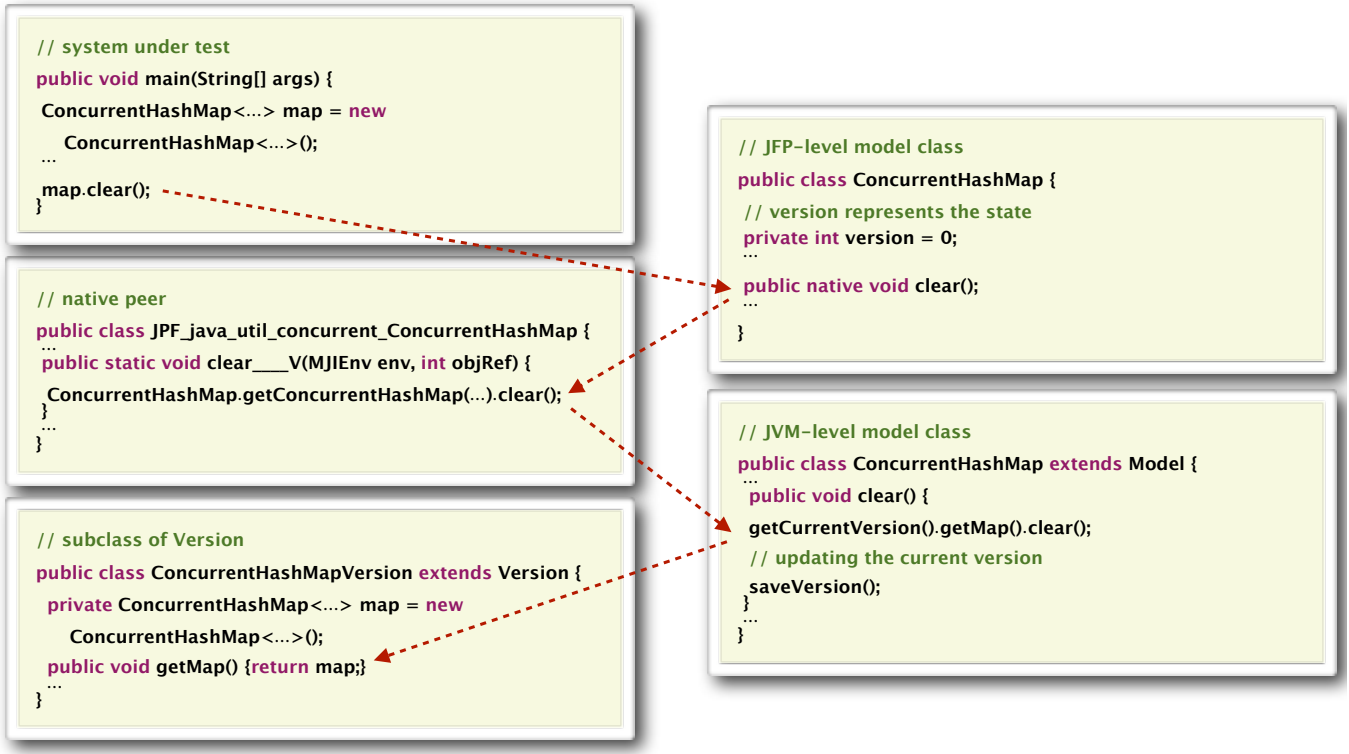

Fig. 4. Transfering execution from JPF to the host JVM

In order to transfer the execution from the JPF-level model classes to the JVM-level model classes, we use the MJI feature of JPF. For each JPF-level model class, that represents a class from the Java Concurrency Utilities in JPF, jpf-concurrent includes its corresponding native peer. Any method in the JPF-level model class that can be hidden from JPF is declared as native, and its corresponding method is included in the native peer. Deciding which methods can be declared as native is left to the programmer. In most cases all methods can be declared native. However, there are some methods that are hard to implement on the host JVM level.

The reason is that the way that objects are represented in JPF is completely different from the their representation in the host JVM. JPF uses the class ElementInfo to capture objects and each object is represented by a unique integer which is a reference to the heap allocation. For example, for the case of ConcurrentHashMap<K, V>, in our approach, the actual state of an instance of this class is stored in an instance of ConcurrentHashMap<Integer, Integer>. Consider the method get(Object key) of ConcurrentHashMap. According to the Java standard API, this method returns the value to which the specified key is mapped. More formally, if this map contains a mapping from a key k to a value v such that key.equals(k), then this method returns v. Implementing this method at the JVM level is hard, since the given key and the content of the map are all integers representing JPF objects and the actual JVM objects do not exist at this level.

In general, in our approach, the native peer methods play the role as connectors between the methods of the JPF-level model classes and the methods of the JVM-level model classes. That can be seen from the delegation pattern demonstrated in Figure 4 for the class ConcurrentHashMap. It should be noted that for the cases where a concurrent class method cannot be declared as native, our approach implements the method within the JPF-level model class. To reduce the state space, the body of the method is surrounded between the methods beginAtomic() and endAtomic() of the gov.nasa.jpf.vm.Verify class, which makes the method execution atomic. It needs to be noted that code between these methods cannot use any thread blocking operations.

In order to capture the actual state of the concurrent object, we introduced a class called Version. Similar to the JVM-level model classes for each class representing a concurrent class, there exists a subclass of the Version class. These classes are completely unknown to JPF and they run on top of the host JVM. For example, for the case of a ConcurrentHashMap we have ConcurrentHashMapVersion, which stores the actual map entries for objects created on the JPF level.

For each concurrent object, there is an instance of the JVM-level model class. To associate instances of the JPF-

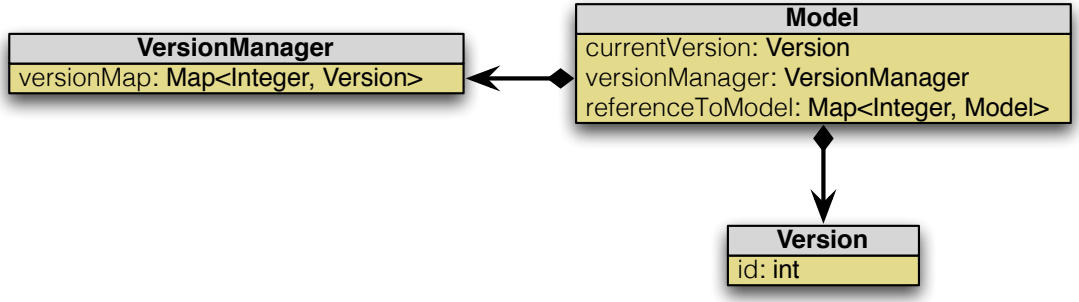

Fig. 5. UML diagram that demonstrates the relationship between main classes of the jpf-concurrent extension

level model classes, that represent concurrent objects in JPF, to instances of JVM-level model classes, that represent concurrent objects in the host JVM, each JVM-level model class contains a static map. This map associates the references of the JPF-level model class to their corresponding JVM-level model classes. This mapping is then used in native peers to retrieve the instance of the correct JVM-level model class.

To each concurrent object, many instances of `Version` can be associated. Each `Version` instance represents a state of the concurrent object that can change during the execution. In JPF-level model classes we have integer field `version` which is used to map current state of the object to `Version` instances on host VM level. To store and update the `Version` instances corresponding to each concurrent object, the class `VersionManager` is introduced, Whenever the state of the concurrent object changes as a result of an operation, `VersionManager` creates a new `Version` object, which stores the updated state and updates `version` variable in JPF-level model class to reflect new version of the object. The UML diagram in Figure 5 demonstrates the relationship between the classes mentioned above.

*Object Removal Process*

The main problem of our approach is that only the objects created on the JPF level can be garbage collected automatically. There is a need for a mechanism to perform the garbage collection on the host JVM level as well. Consider the code snippet from Figure 6.

```
for(int i=0; i<n; i++)
{
    Semaphore s = new Semaphore();
    //interim computation
    s = null;
}
```

Fig. 6. Example of a memory leak in jpf-concurrent

By executing this code, due to `s = null`, instances of the JPF-level model class that correspond to the `Semaphore` objects are created in the JPF level, and they are immediately garbage collected. However, on the host JVM level, n instances of a JVM-level model class that correspond to `Semaphore` objects are created without being garbage collected. Therefore, on the host JVM level, useless objects that correspond to the garbage collected `Semaphore` objects are still occupying the memory.

To avoid this problem, jpf-concurrent implements a listener called `ObjectRemovalListener`. This listener is used to remove the instances of the JVM-level model classes whose corresponding JPF objects have been garbage collected. Every time JPF disposes an object, it sends a notification to `ObjectRemovalListener`. `ObjectRemovalListener` first checks if the class of the object is a concurrent class modelled within jpf-concurrent. If so, it gets rid of an instance of a JVM-level model class that corresponds to the garbage collected object.

## V. BENCHMARKS

All benchmarks have been performed on a Dell PowerEdge R410 machine, with 32GB of RAM, 6 Xeon processors running Fedora 14. Each run has been repeated 10 times. Three classes have been used for benchmarking, `ReentrantLockPerformanceTest`, `ConcurrentHashMapPerformanceTest`, and `AtomicIntegerPerformanceTest`. All of these classes can be found in the jpf-concurrent website [2]. Results have been summarized in Table I.

We have used these classes to show the most prominent types of concurrency constructs that can be found in Java Concurrency Utilities. `ReentrantLock` represents classes that heavily depend on thread blocking instructions. `ConcurrentHashMap` is an example of a data structure that tries to avoid thread blocking in all feasible cases. Finally, `AtomicInteger` represents all classes that use CAS operations [6] to achieve atomicity of instructions.

In all cases, jpf-concurrent shows a speedup. The speedup for the `ReentrantLock` is much higher than for `ConcurrentHashMap` and `AtomicInteger`. This is caused by the fact that using native methods allows us to use fewer thread-blocking operations which are the main source of new transitions, and that can increase the verification time. The smallest speedup has been obtained for `AtomicInteger`, CAS operations are by definition atomic, therefore, moving implementation to native methods did not affect the state space significantly. Due to the exponential increase in test time we were not able to obtain results for a number of threads higher than 6 and 12. Despite this fact, even for small numbers of threads we can see a trend where speedup grows with the number of threads.

[2]http://babelfish.arc.nasa.gov/trac/jpf/wiki/projects/jpf-concurrent

TABLE I
BENCHMARKS OF JPF-CONCURRENT

| Class(Threads) | Without jpf-concurrent | | | With jpf-concurrent | | |
|---|---|---|---|---|---|---|
| | States | Time | Standard deviation | States | Time | Standard deviation |
| ReentrantLock(5) | 620,932 | 00:03:49 | 00:00:08 | 176,775 | 00:00:47 | 00:00:01 |
| ReentrantLock(6) | 9,973,504 | 01:14:32 | 00:01:06 | 1,940,134 | 00:14:10 | 00:00:20 |
| ConcurrentHashMap(11) | 1,054,651 | 00:12:12 | 00:00:12 | 753,727 | 00:05:47 | 00:00:08 |
| ConcurrentHashMap(12) | 4,640,047 | 00:58:22 | 00:00:50 | 3,153,398 | 00:27:10 | 00:00:25 |
| AtomicInteger(11) | 897,075 | 00:06:23 | 00:00:07 | 753,727 | 00:05:14 | 00:00:11 |
| AtomicInteger(12) | 3,747,851 | 00:30:21 | 00:00:38 | 3,153,398 | 00:25:34 | 00:00:24 |

In all cases memory consumption has been a little bit smaller for code run under jpf-concurrent. We attribute this to simplification of data structures that are used on host VM level. Since the difference has been almost negligible, memory usage is not included in the Table I. When used with `ObjectRemovalListener`, our method can produce a significant memory overhead. In the presence of backtracking objects that has been garbage collected on one execution path might be still used on the others. To handle such a situation we need to keep additional information regarding execution paths and monitored objects which results in increased memory usage. Said that, to this point we have not encountered a real life example that would require `ObjectRemovalListener`.

One publicly known case where jpf-concurrent has been used to help verify large NASA applications is [9], but there motivation is support for unimplemented native methods, rather than improved performance.

## VI. CONCLUSIONS

In this paper we proposed a new technique to model Java Concurrency Utilities within JPF. We implemented an extension to JPF called jpf-concurrent which is based on the proposed approach. Our approach not only reduces the state space of the program but also provides support for native calls within the Java Concurrency Utilities that have not been supported in the JPF core project.

Using a set of benchmarks, we have shown that in all cases jpf-concurrent provides a speedup for different types of classes included in Java Concurrency Utilities. At the time of writing this paper the jpf-concurrent project has been actively developed for more than three years, and it is considered one of the most popular extensions of the JPF model checker.

In the future we plan to extend jpf-concurrent to support all classes included in the Java Concurrency Utilities, including new constructs that have been introduced within Java 6 and 7. Another future direction includes supporting operations that rely on time. At this point JPF uses system time which in the presence of backtracking is no longer valid and can lead to bugs that are extremely hard to debug. We also think that researching new methods used for state matching within jpf-concurrent can lead to a significant reduction of the verification time and should be considered in the future.

## ACKNOWLEDGEMENT

The authors would like to thank Dave Parker, Franck van Breugel and anonymous reviewers for their invaluable comments. A special acknowledgement goes to Peter Mehlitz for his constant support during whole development of jpf-concurrent. In 2008 and 2011 development of this project has been funded by Google as a part of Google Summer of Code program.